\documentclass[a4paper, amsfonts, amssymb, amsmath]{article}
\usepackage[utf8]{inputenc}
\usepackage{color}
\usepackage[dvipsnames]{xcolor}
\usepackage{xfrac}
\usepackage{cite}

\usepackage{appendix}
\usepackage{soul}
\usepackage{ulem}
\usepackage{graphicx}
\usepackage{soul}
\usepackage{authblk}
\usepackage{amsmath,amssymb,amsfonts,graphicx}
\usepackage[a4paper,top=2cm,bottom=2cm,left=2cm,right=2cm]{geometry}
\usepackage{tikz}
\usepackage[pdftex, pdftitle={Article}, pdfauthor={Author}]{hyperref}
\usepackage{lscape}

\begin{document}

 \vspace{1.5cm}
\begin{center}
\ \\
{\bf\Large Dark monopoles}

\vspace{0.5cm}

Vicente Vento \vskip 0.7cm

{\it  Departamento de F\'{\i}sica Te\'orica and Instituto de
F\'{\i}sica Corpuscular}

{\it Universidad de Valencia - Consejo Superior de Investigaciones
Cient\'{\i}ficas}

{\it 46100 Burjassot (Val\`encia), Spain, }

{\small Email: vicente.vento@uv.es}

\end{center}
\vskip 1cm \centerline{\bf Abstract}
Monopoles have been a subject of much theoretical and experimental research since they were proposed  to symmetrize Maxwell's equations.
However, no experimental signature of their existence has been detected. Many mechanisms have been proposed to explain this lack of success.
 In here we generalize QED to a two photon theory where the extraordinary photon and the monopole belong to the dark sector, and the ordinary photon 
 and the electron to the non-dark sector. A mixing interaction  transforms both electron and monopole into dyons generating experimental consequences 
 for indirect and direct monopole detection which we analyze.

\vspace{1cm}

\noindent Pacs: 14.80.Hv, 95.30.Cq, 98.70.-f, 98.80.-k


\vspace{1cm}

\section{Introduction}

Dirac's beautiful interpretation of the quantum mechanical behavior of the magnetic monopole led to a technical difficulty associated with the famous Dirac string \cite{Dirac:1931qs,Dirac:1948um}.  The  formalism of  Yang and Wu \cite{Wu:1975es,Wu:1976ge} using fiber bundles and the two photon theory of Cabibbo and  Ferrari \cite{Cabibbo:1962td} simplified the description avoiding strings.  In here we present a model  of two photons \cite{Cabibbo:1962td,Salam:1966bd,Singleton:1996hgp,Scott:2018xgo}, containing an ordinary photon interacting only with electric currents and an extraordinary photon interacting only with magnetic currents, the latter  in the dark sector \cite{Brummer:2009oul,Terning:2018lsv}.  
 We next assume that the dark and ordinary sectors  interact through a portal in a manner that, though feeble, is experimentally accessible. In our portal the interaction takes place because of the kinetic mixing between the dark and the ordinary photons \cite{Holdom:1985ag,Fabbrichesi:2020wbt}. 
Our model differs from that of ref. \cite{Holdom:1985ag} because we work in QED, not in the standard model, to which we add an additional $U(1)$ dark photon \`a la Cabibbo and Ferrari.  In the most general scenario of this model, both  electrons and monopoles become dyons, acquiring electric and magnetic charges.  Based of this result  we propose several experiments to maximize the effect of the new charges, and their observational consequences. A description which has similar motivation to ours is SO(2) dual electrodynamics  of electrically and magnetically charged particles in interaction \cite{Govaerts:2023iqf}. 
 
 \section{Dark photons}
Let us start with a two photon theory whose kinetic part is given by

\begin{equation}
\mathcal{L^\prime}_0 = - \frac{1}{4} \mathcal{F}_{\mu \nu} \mathcal{F}^{\mu \nu} - \frac{1}{4} \mathcal{G}_{\mu \nu} \mathcal{G}^{\mu \nu}\,,
\label{photons}
\end{equation}
where $ \mathcal{F}_{\mu \nu}  = \partial_\mu \mathcal{A}_\nu -\partial_{\nu} \mathcal{A}_\mu $ and  $\mathcal{G}_{\mu \nu}  = \partial_\mu \mathcal{C}_\nu -\partial_{\nu}\mathcal{C}_\mu $.
We  depart from Cabibbo and Ferrari in that we will ascribe the monopoles to  dark matter. Thus, $\mathcal{A}_\mu$ is the ordinary photon that couples only to particles with electric charge and  $\mathcal{C}_\mu$ the extraordinary photon that couples only to particles with magnetic charge. The interaction lagrangian with matter is 
 
 \begin{equation}
 \mathcal{L^\prime}_i = e J^e_\mu \mathcal{A}^\mu + g J^m_\mu \mathcal{C}^\mu\,.
 \label{matter}
 \end{equation}
 where $J^e_\mu $ is the electric current and $J^m_\mu $ the magnetic current. For simplicity we will consider ordinary matter to be made just of electrons and dark matter just of monopoles. We assume  both photons to be massless. In order to generate a portal we assume a kinetic mixing between the photons
 
 \begin{equation}
 \mathcal{L^\prime}_p= -\frac{\varepsilon}{2} \mathcal{F}_{\mu \nu} G^{\mu \nu}\,.
 \end{equation}
 This kinetic term can be diagonalized by transforming the photons \cite{Holdom:1985ag,Fabbrichesi:2020wbt}
 
 \begin{eqnarray}
{ \mathcal{C}^{\mu} \choose  \mathcal{A}^{\mu} } = \left( \begin{array}{cc}{\displaystyle \frac{1}{\sqrt{1-\varepsilon^2}}} &  0 \\
{\displaystyle -\frac{\varepsilon}{\sqrt{1-\varepsilon^2}}} & 1  \end{array} \right)  \left( \begin{array}{cc} \cos \theta & -\sin \theta \\
\sin \theta & \cos \theta  \end{array} \right) 
{ C^{ \mu} \choose  A^{\mu} }  \,,\label{rot}
\end{eqnarray}
 where now we identify $A^\mu$ with the {\it visible} photon and $C^\mu$ with the  {\it dark} photon.  We keep the nomenclature dark photon due to its origin despite the fact that it might couple weakly to ordinary particles as we explain next.
 
 After this transformtion the photon lagrangian becomes diagonal
 
 \begin{equation}
{ \mathcal L}_0 = - \frac{1}{4} F_{\mu \nu} F^{\mu \nu} - \frac{1}{4} G_{\mu \nu} G^{\mu \nu}\,,
\end{equation}
where $ F_{\mu \nu}  = \partial_\mu A_\nu -\partial_{\nu} A_\mu$ and  $G_{\mu \nu}  = \partial_\mu C_\nu -\partial_{\nu} C_\mu.$ 
The interaction lagrangian Eq.(\ref{matter}) becomes

\begin{eqnarray}
\mathcal{L}_i &=& \left( g J^m_\mu\frac{\cos \theta}{\sqrt{1-\varepsilon^2}} + eJ^e_\mu \left( \sin \theta - \frac{\varepsilon \cos \theta}{\sqrt{1-\varepsilon^2}}\right) \right) C^\mu \nonumber \\
& &  + \left( -g J^m_\mu\frac{\sin \theta}{\sqrt{1-\varepsilon^2}} + eJ^e_\mu \left( \cos \theta + \frac{\varepsilon \sin \theta}{\sqrt{1-\varepsilon^2}}\right) \right) A^\mu\,. 
\end{eqnarray}
This equation is very illuminating. It shows that in the transformed theory the electron becomes a dyon with electric and magnetic charge 

\begin{equation}
\left( e \left( \cos \theta + \frac{\varepsilon \sin \theta}{\sqrt{1-\varepsilon^2}}\right) , e \left( \sin \theta - \frac{\varepsilon \cos \theta}{\sqrt{1-\varepsilon^2}}\right)\right)\,,
\label{electron}
\end{equation}
and so does the monopole

\begin{equation}
\left(-g \frac{\sin \theta}{\sqrt{1-\varepsilon^2}},g \frac{\cos \theta}{\sqrt{1-\varepsilon^2}}\right)\,.
\label{monopole}
\end{equation}

Thus the dark photon gets a coupling to the electron and the ordinary photon to the monopole thanks to the mixing term. The new couplings depend not only on the parameter $\varepsilon$, a coupling constant, but also on the parameter $\theta$. For $sin\theta =0$ the ordinary photon must couple to the electron as in ordinary QED. This  fixes the normalization of the fields and therefore different values of $\theta$ lead to different charges, thus to different physical scenarios \cite{Holdom:1985ag,Fabbrichesi:2020wbt}. The aim of the following sections is to discuss various interesting physical scenarios associated with various values for the charges.

 The Schwinger quantization condition for electron-monopole leads to \cite{Schwinger:1975ww}
 
 \begin{equation}
 \frac{e g}{\sqrt{1-\varepsilon^2}} =n ,\;\; n \in N\,.
 \label{schwinger}
 \end{equation}

\section{Experimental consequences}

 From the experimental point of view our approach extends duality to the electron by providing it with a magnetic charge. This is a great experimental advantage if one compares to the search for monopoles. Experiments will deal with a particle easy to produce and easy to control. The handicap is that one expects very small observations of the static magnetic electron charge, thus forcing one to very high precision experiments.
 
\begin{figure}[htb]
\begin{center}
\includegraphics[scale= 0.7]{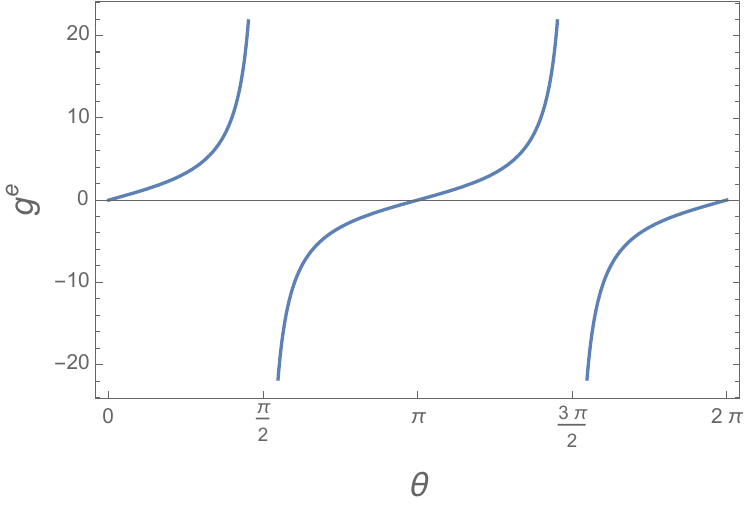} 
\caption{Magnetic charge of the electron as a function of $\theta$ for a small value of $\varepsilon$.}
\label{gelectron}
\end{center}
\end{figure}

 Let us look at the electron charges Eq.(\ref{electron}). It is clear that the mixing parameter $\varepsilon$ has to be very small  as a result of present experimental knowledge. But what about $\theta$? Let us assume that the electric charge of the electron coincides with the extremely well measured experimental charge,
 
\begin{equation}
 e_{exp} = e \left( \cos \theta + \frac{\varepsilon \sin \theta}{\sqrt{1-\varepsilon^2}}\right)\,.
 \end{equation}
 Then for a small value of $\varepsilon$
 
 \begin{equation}
 g^e = e_{exp} \left(\frac{\tan\theta - \frac{\varepsilon}{\sqrt{1-\varepsilon^2}}}{1+ \frac{\varepsilon }{\sqrt{1-\varepsilon^2}}\tan \theta} \right)\,,
 \end{equation}
 which has been plotted in Fig. \ref{gelectron}. The smallness of the magnetic charge effects in the electron imply that the physical region must be  $\theta \sim 0$ .
 
Let us discuss two possibilities satisfying the experimental restriction: 
\begin{center}
i)  $\theta =0$\,,

ii) $\sin\theta=\varepsilon$\,.
\end{center}

 In the first case the electron charges from Eq. (\ref{electron}) become,

 \begin{equation}
( e , -\frac{e \varepsilon}{\sqrt{1-\varepsilon^2}})\,,
\label{electron1}
\end{equation}
while the monopole charges from Eq. (\ref{monopole}) result in

\begin{equation}
 (0, \frac{g}{\sqrt{1-\varepsilon^2}})\,.
\label{monopole1}
\end{equation}

The monopole remains completely in the dark sector, while the electron, couples to the dark photon acquiring a small magnetic charge.
There are many experimental consequences to this result. The most important is to note that the problem of finding magnetic monopoles becomes the problem of measuring a magnetic field produced by a static electron. Despite the enormous difficulty of this endeavor, due to the small size of the effect,  the challenge is not to look for a new particle of unknown mass, but to find a small static property of an abundant well known particle. 

In this case the monopole  cannot  be found by visible  photon scattering, as has been the conventional approach in the past \cite{Epele:2012jn}, since the coupling of monopoles to visible photons vanishes. However, one should remember that the model we are studying here is very naive. Larger values of $\theta$ or more complicated mixing terms, lead  to monopole production via visible photons. However, the important lesson to  be learned from our analysis is that electrons might acquire static magnetic properties by coupling to the dark sector.

Let us now study the second case: $\sin{\theta}= \varepsilon$. Eq. (\ref{electron})  for the electron charges turns into
 \begin{equation}
( \frac{e}{\sqrt{1-\varepsilon^2}},0)\,,
\label{electron2}
\end{equation}
while Eq. (\ref{monopole}) for the monopole charges becomes
\begin{equation}
 ( -\frac{g \varepsilon}{\sqrt{1-\varepsilon^2}}, g)\,.
\label{monopole2}
\end{equation}
This latter electric coupling has been called in the literature mili-charge \cite{Davidson:2000hf}. In this case the experimental observation is far more complicated since we have to deal directly with monopoles whose coupling to visible photons is very weak. Some implications of this result in astrophysics might be possible. In this case the electrons show no special behavior capable of producing clarifying experimental results.

In the intermediate case, $0<\sin \theta<\varepsilon$, we have the whole plethora of phenomena, i.e. electrons with electric and magnetic charge and monopoles with magnetic and electric charge. However, all the observable processes of the electron magnetic charge, or the monopoles electric and magnetic charge will be small.

\section{Experimental consequence of the electron's magnetic charge.}

Let us analyze first some experimental consequences of the electron magnetic charge in the extreme case $\theta =0$, which is an indirect way of providing arguments for the existence of monopoles. An ideal experiment would require large amounts of static electrons gathered in a small region. Since this is not possible, let us study the magnetic field in a circular coil though which  a very intense current  $I = n e v A$ flows. Here, $n$ is the electron density, $A$ the cross area of the coil cable, $v$ the velocity of the electrons in the coil and $e$ the electron electric charge. The ideal scenario will occur in the limit $v \rightarrow 0$. 

\begin{figure}[htb]
\begin{center}
\includegraphics[scale= 0.3, angle=270]{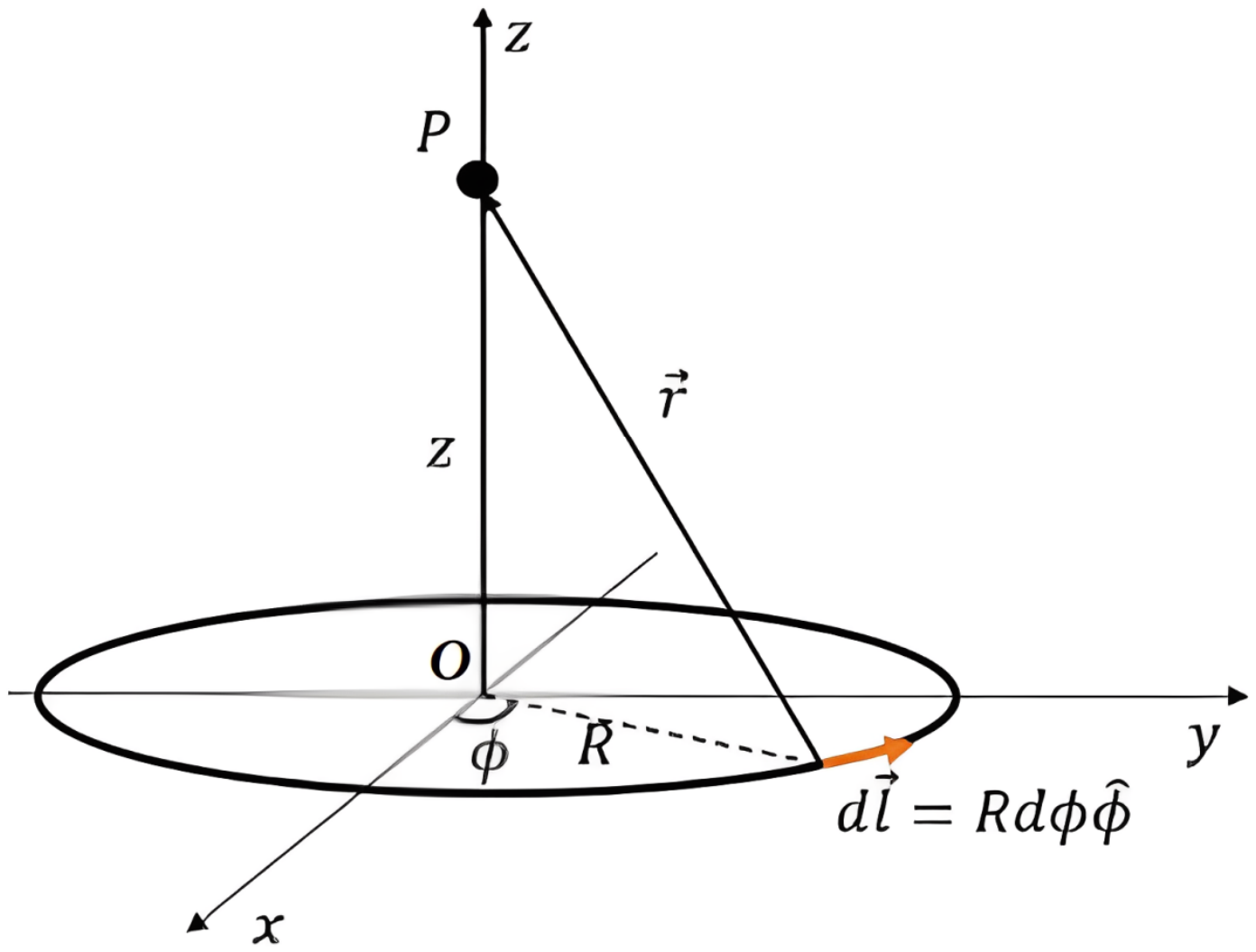} 
\caption{Circular coil of the calculation.}
\label{coil}
\end{center}
\end{figure}

We have a perfect circular coil of radius $R$ and we want to calculate the magnetic field at point $P$ as shown in Fig. \ref{coil}. For the electric current one has to apply the Biot-Savart law \cite{Jackson:1998nia},

\begin{equation}
\vec{B}=  I \oint_l \frac{\vec{dl} \times \hat{r}}{r^2}\,.
\end{equation}
Performing the integral we obtain,

\begin{equation}
\vec{B}_e= 2 \pi \frac{I}{c} \frac{R^2}{(R^2+Z^2)^{\sfrac{3}{2}}} \hat{k}\,,
\end{equation}
where $Z$ is the distance  $\overline{OP}$, $c$ the velocity of light, and $\hat{k}$ the unit vector along the z-axis. 

Let us now calculate the magnetic field associated with the magnetic charge of the electron,

\begin{equation}
\vec{B}_g =  n A g_e \oint_l dl \frac{\hat{r}} {r^2}\,, 
\end{equation}
where  $g_e$ stands here for the magnetic charge of the electron, Eq.(\ref{electron1}). Performing the integral we obtain

\begin{equation}
\vec{B}_g =  2 \pi g_e n A \frac{Z R}{(R^2+Z^2)^{\sfrac{3}{2}}} \hat{k}\,.
\end{equation}
The combined magnetic field in $P$ results in

\begin{equation}
\vec{B}_e+ \vec{B}_g = 2 \pi e n A \frac{R^2}{(R^2+Z^2)^{\sfrac{3}{2}}} \left( \frac{v}{c} - \varepsilon \frac{Z}{R}\right) \hat{k}\,.
\end{equation}

It is apparent that the ideal scenario will occur in the limit $v \rightarrow 0$. The experimental idea one obtains from this analysis is that one must study the variation of the magnetic field on the axis of the coil as a function of the velocity of the electrons and look for a vanishing value for small velocity $v >0$. 

\section{An Aharonov-Bohm experiment}
\begin{figure}[htb]
\begin{center}
\includegraphics[scale= 0.6]{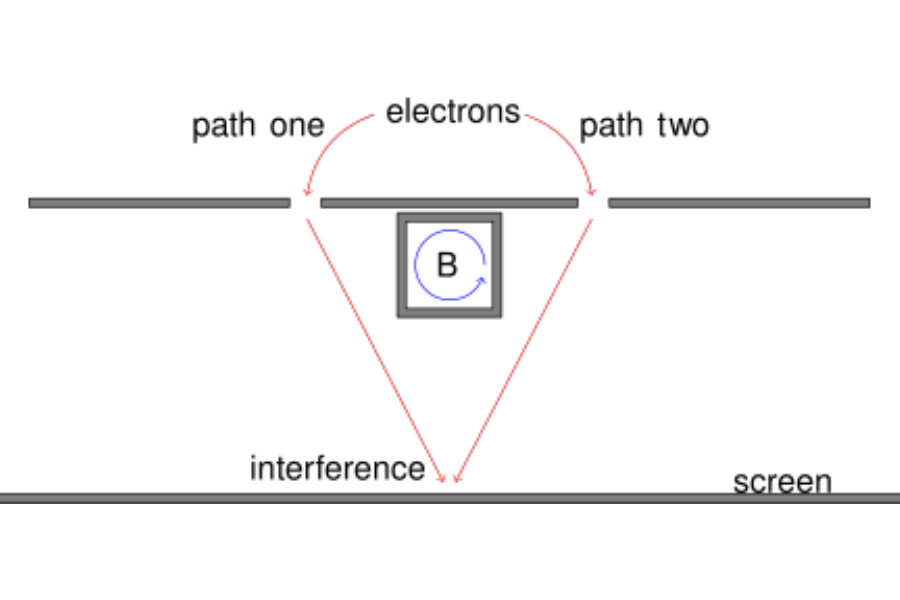} 
\caption{Scheme of double-slit experiment in which the Aharonov-Bohm effect can be observed. The magnetic field can be varied and the interference pattern on the screen of the  electrons which pass through the two slits changes.} 
\label{AB}
\end{center}
\end{figure}

The Aharonov-Bohm effect is a quantum-mechanical phenomenon in which an electrically charged particle is affected by an electromagnetic potential $A_\mu$  despite being confined to a region in which both the magnetic field $\vec{B}$ and electric field $\vec{E}$  are zero \cite{Aharonov:1959fk,Peshkin:1989zz}. The underlying mechanism is the coupling of the electromagnetic potential with the complex phase of a charged particle's wave function, and the Aharonov-Bohm effect is accordingly illustrated by interference experiments \cite{Batelaan:2009zz}. Therefore, particles of electric charge $q$, with the same start and end points, but traveling along two different routes will acquire a phase difference $\Delta \varphi$ determined by the magnetic flux $\Phi_B$ through the area between the paths and given by

\begin{equation}
\Delta \varphi_B = \frac{ q \Phi_B}{\hbar}.
\label{ABeq}
\end{equation}
This phase appears as a consequence of $\vec{B} = \vec{\nabla} \times \vec{A}$. This phase difference can be observed by placing a solenoid between the slits of a double-slit experiment  as shown schematically in Fig,\ref{AB}. The interference pattern shifts when the magnetic field $\vec{B}$ is changed as described by Eq.(\ref{ABeq}),\cite{Batelaan:2009zz}.

Just as the phase of the wave function depends upon the magnetic vector potential, it also depends upon the scalar electric potential, in this case the phase depends on time \cite{VanOudenaarden:1998zz},

\begin{equation}
\Delta \varphi_E = \frac{- q A_0 t}{\hbar},
\end{equation}
which arises from the potential energy $e A_0$.

In the case of the two photon theory the relation between the electromagnetic fields and the electromagnetic potentials change  for static fields to

\begin{align}
\vec{B}= g C_0 + q \vec{\nabla} \times \vec{A}, \\
\vec{E}= q A_0 + g \vec{\nabla} \times \vec{C}.
\end{align}
where $(q,g)$ represent the electric and magnetic charges. Thus in this case the Ahoronov-Bohm effect becomes in terms of the scalar potentials and electric and magnetic fluxes

\begin{align}
\Delta \varphi_B = \frac{- g C_0 t}{\hbar} + \frac{ q \Phi_B}{\hbar}, \label{BB}\\
\Delta \varphi_E = \frac{- q A_0 t}{\hbar} + \frac{ g \Phi_E}{\hbar} \label{EE}. 
\end{align}

Let us analyze the most extreme case for the electron namely $\theta = 0$,  here $q\sim e$ and $g \sim e\varepsilon$. The first equation Eq(\ref{BB}) shows that the time dependence is  crucial for the experimental determination of the magnetic charge of the electron. Keeping the potentials constant and moving the screen of the two slit experiment one obtains a variation of the interference related to the magnetic charge with time. In the second equation  Eq.(\ref{EE}) the crucial ingredient is the electric flux. Keeping the screen fixed and varying the electric flux the interference will be related to the magnetic charge of the electron.

\section{Production of monopoles}

Let us now describe the production of monopoles. We discuss here a model with an intermediate value for the angle $0< \sin{\theta}<\varepsilon$. The charges of the monopole become those of Eqs.(\ref{electron}) and (\ref{monopole}). Using the approximations $\sin{\theta} \sim \theta$, $\theta >> \varepsilon^2$ and $\theta>\theta \varepsilon$ the charges become
for the electron

\begin{equation}
(e, e(\theta- \varepsilon)) ,
\end{equation}
and for the monopole

\begin{equation}
(-g \theta, g).
\end{equation}

\begin{figure}[htb]
\begin{center}
\includegraphics[scale= 0.5]{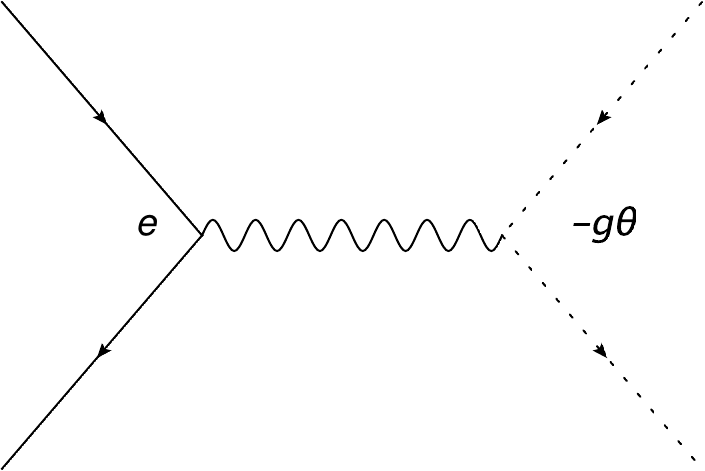} \hspace{1.5cm}\includegraphics[scale= 0.5]{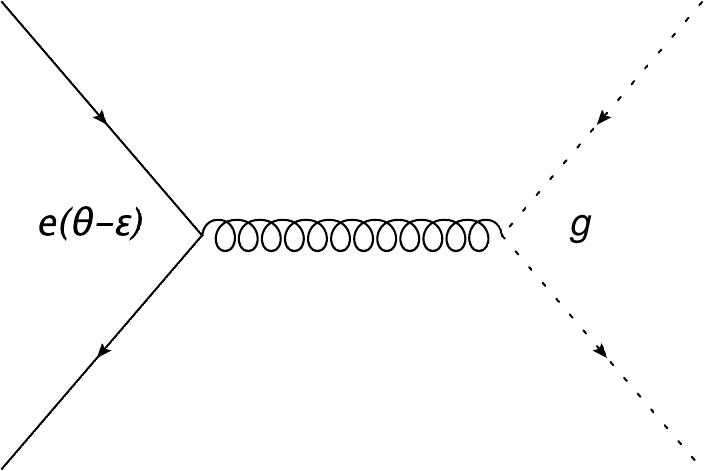} 
\caption{Electron-positron annihilation into monopole-antimonopole. The left figure proceeds via de real photon, while the right figure via the dark photon.}
\label{monopoles}
\end{center}
\end{figure}

One way of producing monopoles  is for example by electron-positron annihilation as shown in  Fig.(\ref{monopoles}). In the left figure we show the production mechanism via a real photon, which decays into monopole-antimonopole via the electric coupling of the monopole. On the right figure we plot the production via a dark photon, which decays via de large magnetic coupling of the monopole. We note that both processes are proportional to $eg$ times a small portal coupling.

Let us show results depending on the spin ($S$) of the monopole and we shall consider the lowest possible charge $\frac{eg}{\sqrt{1-\varepsilon^2}} \sim eg=1$. For $S = 0$ the total cross section is

\begin{equation}
\sigma_{e^+e^-} = \frac{1}{48 \pi} (2\theta-\varepsilon)^2 \frac{(s-4m^2)^{\sfrac{3}{2}}}{s^{\sfrac{5}{2}}},
\label{m0}
\end{equation}
where $\sqrt{s}$ is the center of mass energy and $m$ the monopole mass. For $S=\sfrac{1}{2}$ the total cross section becomes

\begin{equation}
\sigma_{e^+e^-} =  \frac{1}{12 \pi} (2\theta-\varepsilon)^2 \frac{s^{\sfrac{1}{2}} (s-4m^2)^{\sfrac{1}{2}} (s^2+2 s m-6 m^4)}{s^4}.
\label{m0.5}
\end{equation}

\begin{figure}[htb]
\begin{center}
\includegraphics[scale= 0.8]{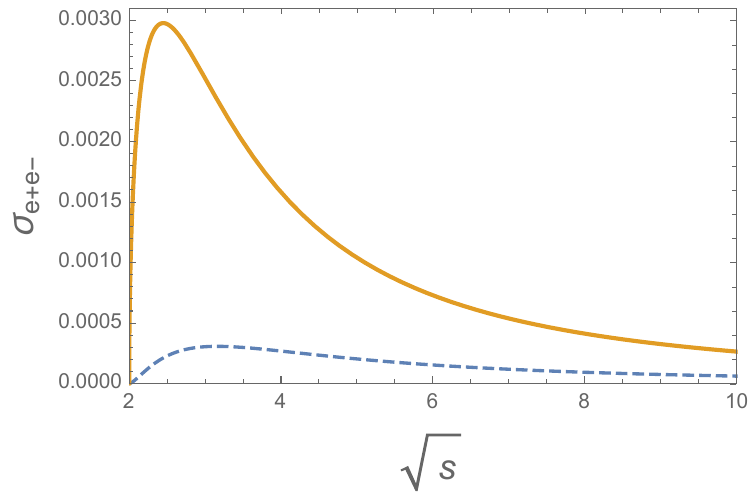} 
\caption{The structure of the total cross sections for $e^+e^-$ annihilation into monopole-antimonople: for $S=0$(dashed) and for $S=\sfrac{1}{2}$ (solid). We use here $(2 \theta-\varepsilon)^2 =1$ and the cross section in units of $m^{-2}$.} 
\label{cross}
\end{center}
\end{figure}

In Fig. \ref{cross} we plot these cross sections for $(2\theta-\varepsilon)^2=1$ in units of $m^{-2}$.

The maximum of the cross sections occur: for $S=0$ at $s= 10 \,m^2$, $\sigma=\sfrac{0.0003082}{m^2}$ and  $\sigma=\sfrac{0.0029779}{m^2}$  at $s= 6 \,m^2$ for $S=\sfrac{1}{2}$.

\begin{figure}[htb]
\begin{center}
\includegraphics[scale= 0.8]{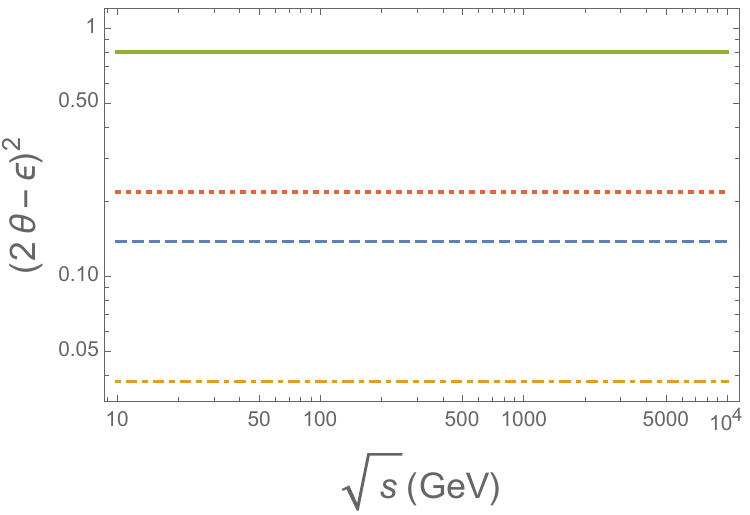} 
\caption{The ratio of the monopole production cross section to $\mu$ pairs (dotted) and to hadrons (dot-dashed)  for $S=0$  monopoles , and  the same  for muon pairs (solid) and to hadrons  (dotted) for $S=\sfrac{1}{2}$  monopoles all ratios multiplied by $10^2$.} 
\label{eemm}
\end{center}
\end{figure}

We next calculate the maximun cross section for monopole masses  $10$ GeV $< m < 10000$ GeV and study the corresponding dependence with $s$. We divide this result by the cross section $e^+e^-  \rightarrow \mu^+\mu^-$,  and by that of  $e^+e^- \rightarrow hadrons$. The agreement between theory and experiment for $e^+e^- \rightarrow \mu^+\mu^-$ is excellent ($<1 $\% level) across a wide range of energies, validating QED and the Standard model \cite{ALEPH:2005ab,Delphi:2005ab,RPP:2006ab,RPP:2025ab}. For $e^+e^-\rightarrow hadrons$ the agreement is excellent  in the high energy region ($<1$\% level); the agrement is good  in the intermediate-energy range ($<5$\% level) and poorer agreement is reached in the low energy region limited by non-perturbative QCD and hadronic resonance modeling \cite{RPP:2006ab,RPP:2025ab}.  In our case we are interested in the high energy region where the experimental precission to produce monopoles has to be at the maximum at the $1$\% level. By using this limit we find upper bounds for the coupling constant appearing  in Eq.(\ref{m0}) and Eq.(\ref{m0.5}), $(2 \theta - \varepsilon)^2$, which we  show in Fig.\ref{eemm}. The extremely good agreement for these  cross sections in the high energy region between experiment and theory restricts greatly the region of the couplings. All the cross sections entering the analysis have been calculated to lowest order, that is the reason behind the absence of $s$ dependence in the ratios. 

\begin{figure}[htb]
\begin{center}
\includegraphics[scale= 0.6]{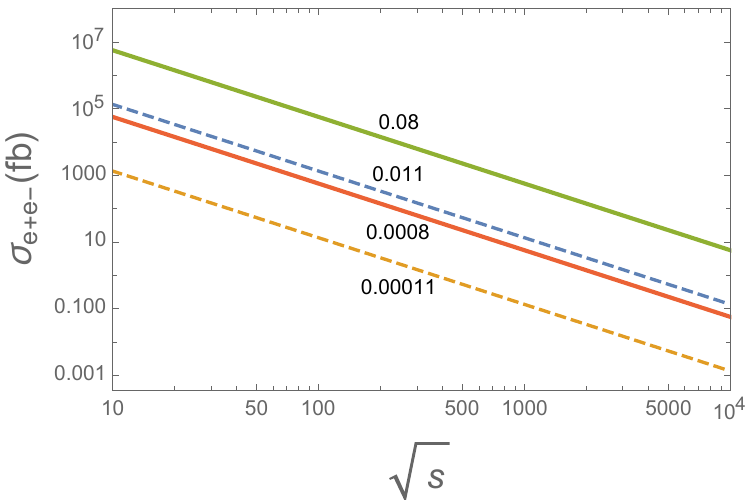} \hspace{1cm}\includegraphics[scale= 0.6]{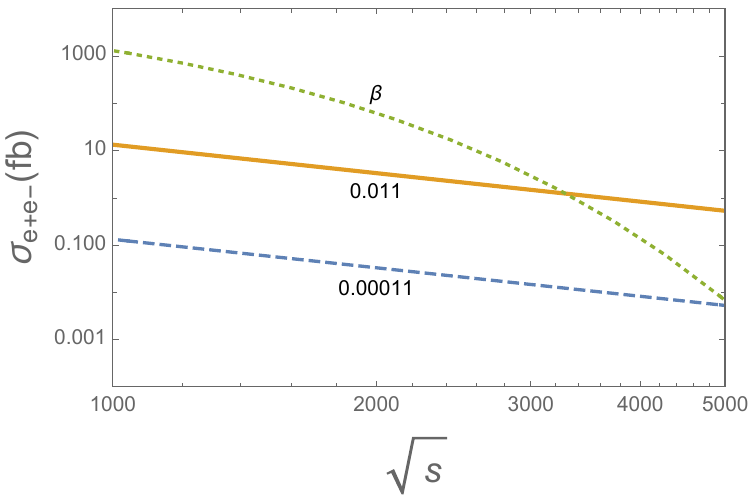}
\caption{The left figure shows the  monopole production cross section for the shown couplings both for $S=0$  monopoles (dashed) , and   for $S=\sfrac{1}{2}$  monopoles (solid). The right figure contains the maximum scalar cross section for couplings 0.011 (solid) and 0.00011 (dashed) in our model compared with the production cross section of ref. \cite{Epele:2012jn} with $\beta$ coupling (dotted)  in the same energy range.}
\label{maxcross}
\end{center}
\end{figure}

Next we plot in the left hand side of Fig. \ref{maxcross}  the maximum monopole production cross section as a function of center of mass energy for 0.1 and 0.001 times the limiting coupling constants, both for the $S=0$ case (dashed) and for the $S=\sfrac{1}{2}$  (solid). Recall that the maximum cross section occurs for $s=10 m^2$ in the scalar case and $s=6m^2$ in the fermion case (see Fig. \ref{cross}), thus each value of the center-of-mass energy has an associated monopole mass. In these ranges of couplings the effect is small but measurable, similar to that obtained by calculations in other schemes, as we show in the right hand side of Fig. \ref{maxcross}, where we compare the present results with those of the calculation of ref. \cite{Epele:2012jn} for the same energy range. It is immediately apparent from our results that the mass of the monopole plays a crucial role. The smaller the mass the larger the detection cross section will be. 

There is a huge difference between the previous ordinary matter calculations and the present one, namely in previous analysis the monopole couples with electromagnetic strength to ordinary photons, and therefore the monopoles can be detected by  their radiation to ordinary photons or the monopole-antimonopole pairs can be detected by their annihilation into ordinary photons \cite{Epele:2012jn,Fanchiotti:2017nkk}, this is not the case here, since the monopole and the monopole-antimonopole pairs couple weakly to ordinary photons. Thus our calculation favors direct detection of monopoles (antimonopoles) \cite{MoEDAL:2014ttp} over the detection via photon radiation.

\section{Conclusions}

We have described an electromagnetic theory for electrons and monopoles with two photons. We assume the magnetic sector of the theory to be associated with a dark sector. We propose a naive coupling between the two photons and diagonalize the interaction in order to separate the ordinary photon from the dark photon. In this process the electron acquires a magnetic charge and the monopole an electric charge. We have analyzed two limiting scenarios characterized by the value of the parameter $\theta$. In the case $\theta = 0$ the monopole remains a dark monopole decoupled from the real world, while the electron acquires a very small magnetic charge. This magnetic charge inspires experimental ways of investigating the ideas behind this approach. We have performed two analysis to see how the magnetic charge of the electron could be determined. The first experimental proposal is a generalization of a Biot-Savart experiment, while the second generalizes the Aharonov-Bohm experiment. These two proposals benefit from the peculiarity of the electron in this model, which becomes a dyon with a very tiny magnetic charge.

In the other scenario $\sin \theta = \varepsilon$ the monopole acquires a very small electric charge, and  therefore production in accelerators is out of the question. Some implications of our results in astrophysics might be possible. If we relax the latter coupling accepting a smaller value of the angle, we have a model with both electric and magnetic charges for electron and monopoles. In this case the two previous experimental proposals are also feasible, and moreover, we can produce monopoles. The production cross section is proportional to $e^2 g^2 $ times a small portal coupling, either $\theta$ or $\theta-\varepsilon$, therefore the model predicts a small production cross section. This might be the reason for not having been able to detect monopoles. We have studied the possible upper limits of the couplings by comparing the production by electron-positron annihilation of  monopole-antimonopole pairs to  $\mu^+\mu^-$  pairs (or hadrons) below the limit of precision of the latter. We see that these upper limits although small are not completely negligible. However, the fact that dark monopoles couple weakly to ordinary photons makes their detection via their decay into ordinary photons further suppressed. Direct detection of monopoles is favored over detection via their radiation.

\section*{Acknowledgments}
This work was inspired by conversations with Huner Fanchiotti and Carlos Garc\'{\i}a Canal. VV acknowledges illuminating email exchanges  with D. Singleton regarding the two photon model, Jan Govaerts regarding the SO(2) model, Marco Fabbrichesi regarding some aspects of ref. \cite{Fabbrichesi:2020wbt}, and Arcadi Santamaria regarding field redefinitions. VV would like to thank Lukas Theussl for help with JaxoDraw \cite{Binosi:2008ig}. Support by Grant CNS2022-135688 Consolidaci\'on Investigadora is acknowledged.

\section*{Data Availability Statement} 
This manuscript has no associated data since all the data generated appear in the figures.

\end{document}